# EARTH'S HEAT SOURCE - THE SUN

**Oliver K. Manuel**

*Emeritus Professor, Space and Nuclear Studies*
*University of Missouri, Rolla, MO 65401*
*Associate, Climate & Solar Science Institute*
*625 Broadway, Cape Girardeau, MO 63701*
*E-mail: omatumr@yahoo.com*
*Website: http://www.omatumr.com*

**ABSTRACT**
The Sun encompasses planet Earth, supplies the heat that warms it, and even shakes it. The United Nation's Intergovernmental Panel on Climate Change (IPCC) assumed that solar influence on Earth's climate is limited to changes in solar irradiance and adopted the consensus opinion of a hydrogen-filled Sun—the Standard Solar Model (SSM). They did not consider the alternative solar model and instead adopted another consensus opinion: Anthropogenic greenhouse gases play a dominant role in climate change. The SSM fails to explain the solar wind, solar cycles, and the empirical link of solar surface activity with Earth's changing climate. The alternative solar model—molded from an embarrassingly large number of unexpected observations that space-age measurements revealed since 1959—explains not only these puzzles but also how closely linked interactions between the Sun and its planets and other celestial bodies induce turbulent cycles of secondary solar characteristics that significantly affect Earth's climate.

## 1. INTRODUCTION

Much of what we hear from the scientific community concerning AGW (Anthropogenic Global Warming) is based on an obsolete model of the Sun, a misunderstanding of the ways that Earth is connected to this unstable heat source, and on politically driven conclusions that come either directly or indirectly from the UN's Intergovernmental Panel on Climate Change (IPCC). The IPCC admits that it *"does not conduct any research nor does it monitor climate related data or parameters"* [1].

Instead, IPCC operates under a myopic mandate to assess *"the risk of human-induced climate change"* [1]. Not surprisingly, the IPCC found what they were looking for.

This paper is concerned with fundamental flaws in the currently fashionable model of Earth's heat source—the Sun. Errors in theoretical models may significantly impact us, as Alan Greenspan recently conceded [2] when the world economy started to crumble.

Statements in the IPCC Summary Report for Policymakers [3] and the Assessment Report of IPCC's Working Group I [4] seem to contradict the claim that IPCC reports are *". . neutral with respect to policy"* [1]. The reports preserve the illusion that man is primarily responsible for the Earth's current warm period by discounting sources of heating and cooling that do not fit into this narrow view with factually soft, inexact statements like these: *"Most of the observed increase in global average temperatures is very likely due to the observed increase in anthropogenic GHG (Green House Gas) concentrations"* [3]. *"During the past 50 years, the sum of solar and volcanic forcings would likely have produced cooling"* [3]. *"Changes in solar irradiance since 1750 are estimated to cause a radiative forcing of +0.12 [+0.06 to +0.30] $W\,m^{-2}$ . . ."* [4]. Thus, the IPCC concludes that what man has caused, man can now remedy. That misconception is a greater danger to us than the illusion of anthropogenic global warming. Were it not for these remedies, whether warming was anthropogenic or natural, it would be purely academic. It is not.

The vision of our stormy Sun as a mild–mannered, hydrogen fusion furnace—the Standard Solar Model, or SSM—is basic to this misunderstanding. This interpretation of Earth's source of warmth ignores electromagnetic links [5, 6] and secondary effects of the Sun on our climate—cycles of solar eruptions, cosmic rays, sunspots and changes in magnetic polarity and intensity that follow gravitational interactions of the Sun's dense, energetic core with planets and galactic objects [7-17 and references therein]. Analyses of planets, the Moon, the solar wind, solar flares, the solar photosphere, and ordinary meteorites show that our Sun is actually the violent, ill-mannered remains of a supernova that once ejected all of the heavier elements on Earth and in the solar system and now selectively moves lightweight elements into a veneer of H and He that covers the Sun's energetic neutron core [18]. This brings the IPCC conclusions into question and, more importantly, the draconian solutions that some policymakers advocate.

## 2. THE COMPOSITION, ORIGIN AND OPERATION OF THE SUN

The Sun and the Earth are intimately connected. These objects are mistakenly perceived as separate entities, in large part because visible light from the photosphere produces the illusion of a solar "surface" between the Earth and the Sun. The TRACE spacecraft recorded images of rigid, iron-rich features (Figure 3) beneath the fluid photosphere [18, 19]. Layers above the fluid solar "surface" are: a.) First the chromosphere; b.) Then, the corona; and c.) Finally the heliosphere extends beyond the planets to ~100 AU (astronomical units) from the Sun [20]. Earth and the other planets glide through the heliosphere and are almost instantly affected by heliospheric disturbances. That was the main lesson that Richard Carrington learned in 1859 [21]. World communications were interrupted as telegraph systems crashed and Earth was engulfed in a blood-red aurora following a massive solar eruption that Carrington recorded at 11:18 am on September 1, 1859. NASA has confirmed that Earth's orbit lies inside the heliosphere, an invisible sheath of solar wind particles (mostly $H^+$ and $e^-$ ions) and the solar magnetic field that extends more than 100 AU (astronomical units) above the visible "surface" of the Sun [20]. Recently it was announced that vibrations from the Sun even shake our planet [22].

**2.1 The Composition of the Solar Photosphere**

The Standard Solar Model claims that the chemical composition of the interior of the Sun is essentially the same as that shown below (Figure 1) for its visible "surface", the photosphere [23]:

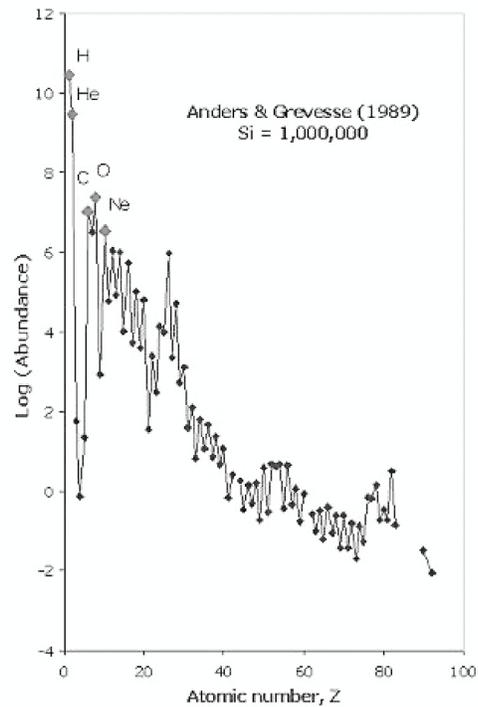

**Figure 1**. *Solar abundances of the elements commonly refer to their abundances in the photosphere [23]. Large diamonds identify the five most abundant elements in the photosphere: H, He, C, O and Ne. These abundances generally decline exponentially with atomic number (Z), except for a deficit of elements with unusually low nuclear stability (Li, Be and B) and an excess of elements with unusually high nuclear stability (Fe).*

**2.2 Solar Mass Fractionation and Bulk Composition**
Two measurements [24, 25] independently confirmed that the Sun selectively moves lightweight atoms upward to generate the mass-fractionated veneer of elements (91% H and 9% He) shown in Figure 1. The left side of Figure 2 shows the mass fractionation pattern seen across 22 isotopes of noble gases in the solar wind [24]. The right side shows the mass fractionation pattern seen across 72 s-products in the solar photosphere [25]. (S-products are atoms made in red giant stars by slow neutron capture [26]).

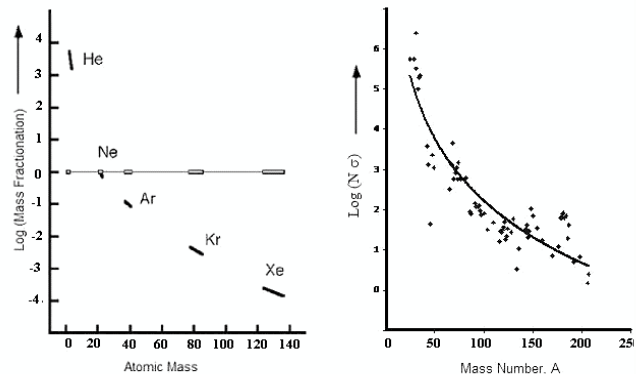

**Figure 2**. *On the left, 22 noble gas isotopes in the solar wind (filled bars) are mass fractionated relative to planetary noble gases (open bars) [24]. This mass fractionation is recorded across a mass range from 3 to 136 mass units (mu). On the right, s-products in the photosphere are mass fractionated relative to the values expected from slow neutron capture [26]. This mass fractionation is recorded across the abundances of 72 s-products in the photosphere, spanning a mass range from 25 to 207 mu [25].*

The five more abundant elements in the interior of the Sun are revealed to be Fe, O, Si, Ni, and S when chemical components of the photosphere (Figure 1) are corrected for the empirical mass fractionation shown on either the left or right sides of Figure 2.

Figure 3 (top of next page) shows the abundances of elements in the interior of the bulk Sun when elements in the photosphere (Figure 1) are corrected for the empirical mass fractionation that was observed across isotopes in the solar wind (left side of Figure 2).

**2.3 The Puzzling Interior of the Sun**
The <u>average</u> solar density does not falsify the analytical results shown in Figures 2 and 3. In fact, the probability is essentially zero ($P < 2 \times 10^{-33}$) [28] that the mass fractionation seen across isotopes in the solar wind would fortuitously identify the more abundant elements in meteorites [27] as the ones that are also more abundant in the Sun [24]. However, the <u>average</u> solar density and many other observations show that the internal structure of the Sun is indeed complicated. The Sun vibrates like a pulsar [29] and has rigid, iron-rich features beneath the photosphere [19]. G-waves from the solar core literally shake the planet Earth [22].

Densities within the Sun span many orders of magnitude. The average overall density of the Sun, which depends on both internal structure and composition, may be as meaningless as the average overall density in the Rutherford-Bohr model of the atom. More than that we cannot say, except that the internal structure of the Sun is unknown and appears to be very complex.

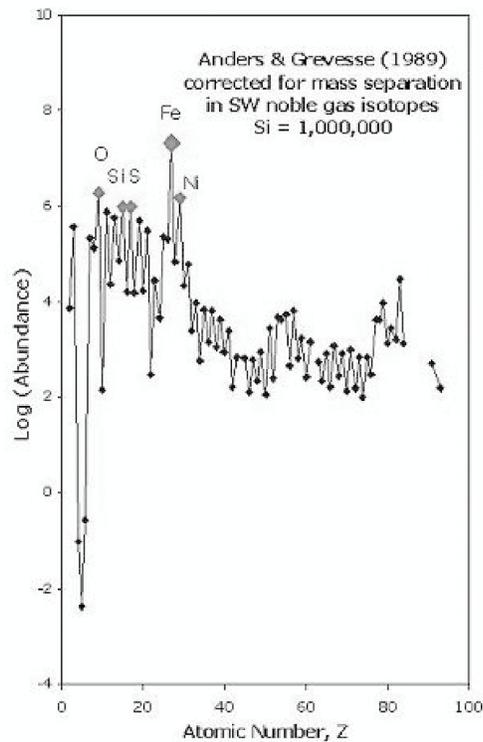

**Figure 3.** *Large diamonds identify Fe, O, Ni, Si and S as the five more abundant elements in the interior of the Sun, calculated by correcting the abundance of elements in the photosphere [23] for the mass fractionation observed across the isotopes of elements in the solar wind [24]. These are also the five more abundant elements in meteorites [27].*

The solar density is about $10^{-6}$ g/cm$^3$ in the photosphere, and the density of material in the nuclear solar core is at least equal to that of the atomic nucleus, $\sim 10^{+15}$ g/cm$^3$. The Sun extends outward about 100 AU above the photosphere and has an average overall density of about $1.4 \times 10^{-13}$ g/cm$^3$. The density of material above the photosphere has a wave-like structure, with wave crests at the orbit of each of the planets in the equatorial plane, 0.39 AU, 0.72 AU, 1.00 AU, . . . 30 AU. The structure is also complex beneath the photosphere.

The total mass of material beneath the visible solar "surface" has an <u>average</u> density of about 1.4 g/cm$^3$. Niels Bohr [30] noted similarities between the structure of the atom and that of the solar system. Others [31, 32] suggested that the internal structure of stars mimics that of the atomic nucleus. Carl Rouse [33] reported evidence of an iron-rich solar core in 1985, and Michael Mozina [19] noticed rigid, iron-rich solar structures closer to the solar "surface" in 2005 (Figure 4). Lockheed Martin made this "running difference" image of an active region of the Sun (AR 9143) from photographs that the TRACE satellite took, using a 171 Å filter to enhance light emissions from iron [18,19].

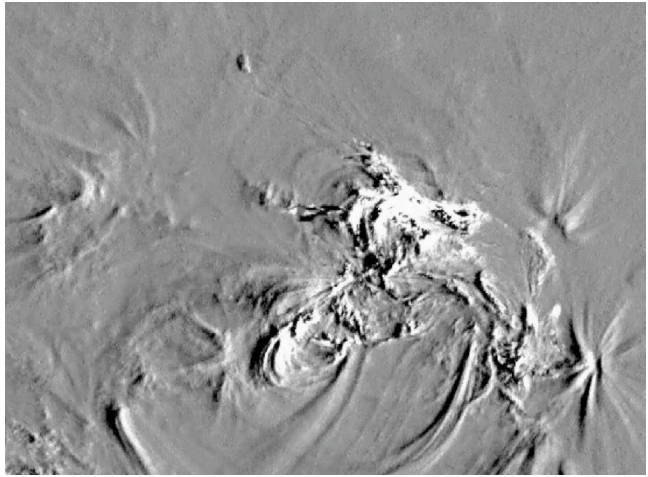

**Figure 4.** *This is a "running difference" image of an active region of the Sun (AR 9143) that the TRACE satellite made on 28 August 2000 using a 171 Å filter. This filter is specifically sensitive to light emissions from iron ions, Fe IX and Fe X. Lockheed Martin made a movie of "running difference" images that the TRACE satellite recorded during the C3.3 solar flare and mass ejection from this same region. To view the video recording of this flare event, go to:*
*http://trace.lmsal.com/POD/movies/T171_000828.avi*

**2.4 The Origin of the Solar System**
The Sun and the Earth were born together from the debris of a violent stellar explosion [18] about five billion years ago [34].

The first hint that a supernova, rather than an interstellar cloud, made the solar system—the Sun and the planets, moons, comets, asteroids and other rubble that orbits it—came in 1960, when John Reynolds reported the decay product of extinct $^{129}$I (16 My half-life) and strange abundances of the nine stable isotopes of primordial Xenon in meteorites [43, 44] and Paul Kuroda reported the decay product of extinct $^{244}$Pu (82 My half-life) in air [45]. These discoveries were difficult to reconcile with the idea that the solar system formed out of an interstellar cloud.

Fowler, Greenstein and Hoyle [46] suggested that Deuterium, Li, Be, B, $^{129}$I and other short-lived radioactive nuclei were produced in the solar system itself. Rapid neutron capture, the r-process, in a supernova is the only known way to make $^{244}$Pu in stars [26]. Figure 5 shows a possible scenario for making $^{244}$Pu and other short-lived isotopes in the early solar system. Numerous isotope and element analyses of meteorites, the Moon, and other planets since 1959 leave little doubt that the solar system formed out of fresh supernova debris (See ref. [18, 34-45] and references cited therein).

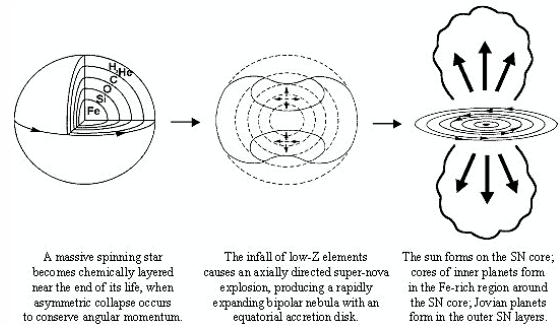

*Figure 5. This is a schematic drawing of the violent birth of the solar system from a supernova. This conclusion about the origin of the solar system is based on numerous reports of the decay products of short-lived nuclides and linked chemical and isotopic heterogeneities from stellar nuclear reactions in the material that formed the Earth, the Sun, the meteorites and other planets [18, 34-45 and hundreds of references therein].*

**2.5 The Sun's Source of Heat**

Abundant elements in the Sun have high nuclear stability and cannot be the source of solar luminosity, solar neutrinos and solar-wind hydrogen that pour from the solar surface. These puzzling features of the Sun remained a mystery until 2000.

That year five graduate students in an advanced nuclear science class at the University of Missouri-Rolla (Chem. 471)—Cynthia Bolon, Shelonda Finch, Daniel Ragland, Matthew Seelke and Bing Zhang—helped construct a 3-D plot of reduced nuclear variables, M/A (mass per nucleon, or potential energy per nucleon) and Z/A (charge density, or charge per nucleon), for each of the 3,000 known nuclides in the ground state [47]. The results were first published as the "Cradle of the Nuclides" on the cover of the book, *"Origin of Elements in the Solar System: Implications of Post 1957 Observations"* [41] and then later elsewhere [18, 48, 49].

Raw data [47] for the "Cradle of the Nuclides" are shown below, on the left side of Figure 6. On the right side of Figure 6 are mass parabolas, defined by the data points at each mass number, A > 1, and the intersections of those empirically–defined mass parabolas with the back plane (at Z/A = 1) and with the front plane at Z/A (at Z/A =0).

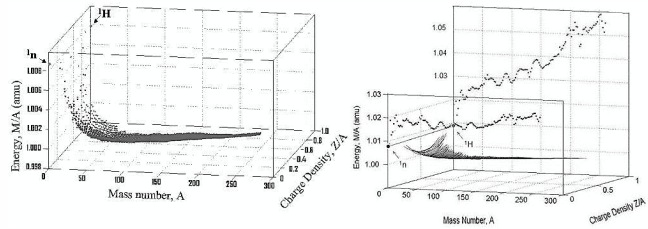

**Figure 6.** *The "Cradle of the Nuclides" on the left shows the potential energy per nucleon for all stable and radioactive nuclides that were known in 2000 [47]. The more stable nuclides have lower values of M/A and occupy lower positions in the cradle. Nuclei that are radioactive or readily consumed by fusion or fission occupy higher positions. In the figure on the right, mass parabolas through data points at each value of A>1 intersect the front plane at { Z/A = 0, M/A = (M/A)$_{neutron}$ + ~10 MeV }.*

The right side of Figure 6 shows the mass parabolas defined by the data at each value of A>1 [47]. Intersections of mass parabolas with the front plane at Z/A = 0 show the potential energy per nucleon, M/A, for assemblages of pure neutrons. Intersections of mass parabolas with the back plane at Z/A = 1 show the potential energy per nucleon, M/A, for assemblages of pure protons. Repulsion between positive charges causes the value of M/A at Z/A = 1 to be larger than that at Z/A =0 [48]. The intercepts at Z/A = 0 and Z/A = 1.0 have peaks and valleys at the same mass numbers because of the potential energy (mass) produced by tight or loose packing of nucleons that repel each other [49].

Systematic properties of nuclei in Figure 6 reveal strong attractive interactions between unlike nucleons (n–p) and symmetric, repulsive interactions between like nucleons (n–n or p–p) that are further increased by Coulomb repulsion between positive charges on protons [48, 49]. The nuclear mass data shown in Figure 6 indicate that neutrons in a neutron star will be in an excited state, with about +10-22 MeV more energy than a free neutron.

Thus, neutron repulsion in the core of the Sun triggers a series of reactions that generate solar luminosity, neutrinos, and an outflow of $3 \times 10^{43}$ H$^+$ per year in the solar wind [18]:

1. <u>Neutron emission</u> from the solar core  ***Generates >57% of solar luminosity***
$$<{}_0^1n> \longrightarrow {}_0^1n \; + \sim 10\text{-}22 \text{ MeV}$$

2. <u>Neutron decay</u>  ***Generates < 5% of solar luminosity***
$$_0^1n \longrightarrow {}_1^1H^+ + e^- + \text{anti-}\nu \; + 0.78 \text{ MeV}$$

3. <u>Fusion</u> and upward migration of H+  ***Generates <38% of solar luminosity***
$$4 \; _1^1H^+ + 2 \; e^- \longrightarrow {}_2^4He^{++} + 2 \; \nu \; + 27 \text{ MeV}$$

4. <u>Escape of excess H$^+$</u> in the solar wind  ***Generates 100% of solar-wind H***
$$3 \times 10^{43} \text{ H}^+/\text{yr} \longrightarrow \text{Departs in solar wind}$$

**2.6 The Link of Earth's Climate with the Iron Sun**

Several papers in this special volume of *Energy & Environment* and in the earlier scientific literature empirically verify the connection of Earth's climate with cyclic changes in solar activity and with changes in solar inertial motion [7-17]. There is no doubt that sunspot production is linked to orbital motion of the planets and to velocity changes in the Sun, as it is jerked-like a yo–yo on a string–about the constantly changing centre-of-mass (barycentre) of the solar system. However, the mechanism for sunspot production has remained hidden. The reason for this empirical fact would remain a mystery [14-16] if the Sun were in fact the homogeneous object described by the Standard Solar Model. It certainly is not.

The Sun is stratified, covered with a surface veneer of hydrogen–the most lightweight of all elements–and centered on an energetic core of extreme nuclear density. The depth of this dense, energetic solar core shifts relative to the solar "surface" as gravitational forces cause the Sun to experience abrupt acceleration and deceleration in its orbit about the barycentre of the solar system. The Sun's resulting irregular orbit is shown in Figure 7.

In 2007 Alexander et al. [15] noted, *"There can be no doubt that it is the influence of the changing positions of the major planets that is the direct cause of sunspot activity. The actual mechanism for sunspot production as a result of galactic velocity changes in the sun has yet to be determined"*. The Sun's dense, energetic core of neutrons explains this mystery.

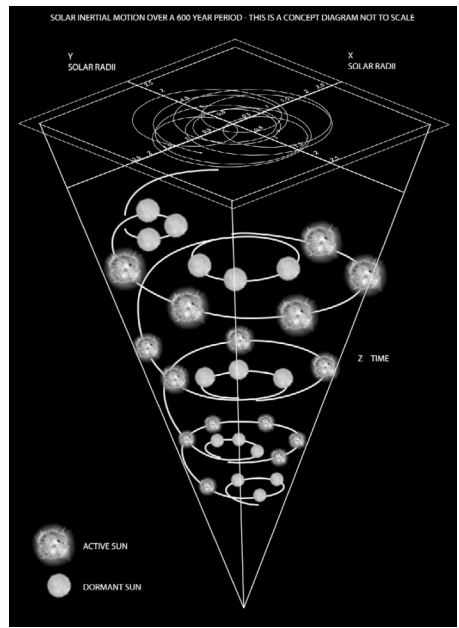

**Figure 7.** *This schematic drawing illustrates sudden shifts in the solar inertial motion (SIM) as the Sun travels in an epitrochiod-shaped orbit about the centre-of-mass of the solar system. This 2006 drawing by Daniel Brunato, University of Canberra, was reproduced by permission from Richard Mackey [16]. Shown here are three complete orbits of the Sun, each of which takes about 179 years. Each solar orbit consists of about eight, 22-year solar cycles [16]. The total time span shown in Figure 7 is therefore three 179-year solar cycles [8], or about 600 years.*

**2.7 Politics, Science and the IPCC**
One need not know all of the ways that Earth is connected to the Sun to be able to say with certainty that the IPCC was wrong to assume that solar irradiance is the only solar variable that produces changes in Earth's climate. As this paper was being completed, Dr. David Sibeck reported the discovery of yet another link between the Sun and the Earth [50].

Dr. Ralph J. Cicerone, President of the US National Academy of Sciences, led the 2001 NAS study of climate change that persuaded US President George Bush to support the IPCC. Six years later Dr. Frederick Seitz, the distinguished former NAS President, replied in the foreword to the 2007 NIPCC Report [51], *". . . we do not currently have any convincing evidence or observations of significant climate change from other than natural causes."*

Why is the scientific community so divided on an issue that can only be resolved by detached observations? Is it perhaps because the IPCC has been less than completely policy-neutral with regard to its findings? They claim to be, but there is now significant evidence that politicians rather than scientists control the tone and content of the final reports. Actually, everything goes quite well in the report process through the second draft, following the first round of comments. However, after the Summary for Policy Makers is published, *"Changes (other than grammatical or minor editorial changes) made after acceptance by the Working Group or the Panel shall be those necessary to ensure consistency with the Summary for Policymakers or the Overview Chapter"* [52]. In other words, the scientific reports that have been reviewed by some of the world's finest minds are altered to agree with a politically generated summary document.

So, do politicians affect the IPCC process outcome? The UN's IPCC falsely implies that 2500 scientists endorse the full AR4 report [53] when only a small percentage of the reviewers made comment on multiple chapters and a majority (58.1%) of negative comments on the critical Chapter 9, *"Understanding and Attributing Climate Change,"* were rejected [54]. The Independent Summary for Policy Makers (ISPM) [55] derived from the second draft of the AR4 Working Group1 [53], before its final modification, came to a quite different conclusion. Some of the contributors to ISPM [55] were among the 2500 official IPCC reviewers. Similarly, the NIPCC project, *"Nature, not human activity rules the climate"* [51], came to a like determination using a broader range of literature. The NIPCC scientists [51] did the work for no financial considerations, avoiding criticisms based on profit motives. In short, these two reports [51, 54] cast severe doubts on the conclusions voiced by the IPCC.

**3. CONCLUSIONS**
IPCC reports on the dangers of AGW (Anthropogenic Global Warming) are based on an obsolete model of the Sun, a misunderstanding of the many ways that Earth is connected to its heat source, and on politically driven conclusions. The scientists are not at fault. The die for the present disaster was likely cast in the late 1940s or early 1950s, when federal research agencies like NSF started using the anonymous review system to obtain consensus opinions. Politicians realized that knowledge is power when World War II ended with an explosive and decisive display of success by the Manhattan Project. I have seen the unholy alliance between politics and science grow since my scientific career started in 1960, despite this warning by US President Dwight D. Eisenhower in his 17 January 1961 Farewell Address to the Nation: ***"The prospect of domination of the nation's scholars by Federal employment, project allocations, and the power of money is ever present and is gravely to be regarded"*** [56].

## 4. ACKNOWLEDGMENTS

I am grateful to Kirt Griffin for encouragement and helpful comments on this manuscript, to Richard Mackey for permission to reproduce Figure 7, to the Foundation for Chemical Research, Inc. for permission to reproduce other figures from FCR reports, to former graduate students–Cynthia Bolon, Shelonda Finch, Daniel Ragland, Matthew Seelke and Bing Zhang–who helped develop the "Cradle of the Nuclides" (Figure 6) and expose repulsive interactions between neutrons, and for moral support from my wife, Caroline, children, former students and friends: Hilton Ratcliffe, Stig Friberg, Barry Ninham, E. Calvin Alexander, Jr., Kiril Panov, Golden Hwaung, William Myers, Marcel Pleess, R. Ganapathy, Marvin Herndon, Matt Insall, Phyllis Johnson, Sumeet Kamat, Michael Mozina, Jim Pauley, Joseph Senne, Nirmal K. Shastri, K. Y. (George) Chiou, Jamie Cochran, Gerry D. Reece, Mike Reno, Jim Bolten, Vince Roach, Joe Szabo, Jim Pauley, and the late Professors Hannes Alfven, Ray Bisplinghoff, Eknath Easwaran, Paul K. Kuroda, John H. Reynolds, Carl Rouse, Dwarka D. Sabu, Glenn T. Seaborg, Mitsunobu Tatsumoto, and Gary Thomas.

**REFERENCES**

1) Intergovernmental Panel on Climate Control (IPCC), See the IPCC Mandate in the first paragraph of *About IPCC*, http://www.ipcc.ch/about/index.htm

2) Andrews, E.L., "Greenspan concedes error on regulation," *The New York Times, Economy Section*, 23 October 2008.

3) IPCC, *Climate Change 2007: Synthesis Report, Summary for Policymakers*, Section 2: Causes of Change, (November 2007) page 5
http://www.ipcc.ch/pdf/assessment-report/ar4/syr/ar4_syr_spm.pdf

4) IPCC, *Climate Change 2007: The Physical Science Basis, Summary for Policymakers, Contribution of Working Group I to the Fourth Assessment Report (WGI Fourth Assessment Report* 2007, February 5, 2007) page 5
http://ipcc-wg1.ucar.edu/wg1/docs/WG1AR4_SPM_Approved_05Feb.pdf

5) McCormmach, Russell, "H. A. Lorentz and the electromagnetic view of nature," Isis, 1970, 61(4) 459-497. http://www.jstor.org/pss/229459?cookieSet=1

6) Akasofu, Syun-Ichi, *Exploring the Secrets of the Aurora*, Springer Publishing Co., New York, NY, 2007, 288 pp.

7) Jose, P.D., "Sun's motion and sunspots", *Astron*. J., 1965, 70, 193-200.

8)  Fairbridge, R.W. and Shirley, J.H., "Prolonged minima and the 179-yr cycle of the solar inertial motion," *Solar Physics*, 1987, 110, 191-220.

9)  Svensmark, Henrik, "Cosmic rays and Earth's climate," *Space Science Reviews*, 2000, 1555-1666.

10) Manuel, O.K., Ninham, B.W. and Friberg, S.E., "Superfluidity in the solar interior: Implications for solar eruptions and climate," *Journal of Fusion Energy*, 2002, 21, 193-198. http://arxiv.org/abs/astro-ph/0501441

11) Landscheidt, Theodor, "New Little Ice Age instead of Global Warming?", *Energy & Environment*, 2003, 114 ( 2/3), 327-350.

12) Rozelot, J-P. and Lefebvre, S., "Is it possible to find a solar signature in the current climatic warming?", *Physics and Chemistry of the Earth*, 2006, 31, Issues 1-3, 41-45.

13) Yousef, S., "80-120 yr long-term solar induced effects on the Earth: Past and predictions," *Physics and Chemistry of the Earth*, 2006, 31, Issues 1-3, 113-122.

14) Shirley, J., "Axial rotation, orbital revolution and solar spin-orbit coupling," *Monthly Notices of the Royal Astronomical Society*, 2006, 368, 280-282.

15) Alexander, W.J.R., Bailey, F., Bredenkamp, D.B., vander Merwe, A., and Willemse, N., "Linkages between solar activity, climate predictability and water resource development," *J. South African Institut. Civil Eng.*, 2007, 49, 32-44.

16) Mackey, Richard, "Rhodes Fairbridge and the idea that the solar system regulates the Earth's climate," *Journal of Coastal Research*, 2007, SI 50 (Proceedings of the Ninth International Coastal Symposium, Gold Coast, Australia) pp. 955-968. http://www.griffith.edu.au/conference/ics2007/pdf/ICS176.pdf

17) Mackey, Richard, "The Sun's role in regulating Earth's climate dynamics," *Energy & Environment*, 2009, 20, this issue.

18) Manuel, O., Kamat, S.A. and Mozina, M., "Isotopes tell origin and operation of the Sun," in *AIP Conference Proceedings*, 2006, 822 (*Proceedings of the First Crisis in Cosmology Conference*, Monção, Portugal, 23-25 June 2005) pp. 206-225. http://arxiv.org/pdf/astro-ph/0510001v1

19) Mozina, Michael "The surface of the Sun," April 2005, http://www.thesurfaceofthesun.com/index.html

20) Nemiroff, R.J., "The Sun's heliosphere & heliopause," *Astronomy Picture of the Day*, June 24, 2002, http://apod.nasa.gov/apod/ap020624.html

21) Clark, Stuart, *The Sun Kings: The Unexpected Tragedy of Richard Carrington and the Tale of How Modern Astronomy Began*, Princeton University Press, Princeton, NJ, 2007, 211 pp.

22) Thomson, D.J., Lanzerotti, L.J., Vernon, F.L., Lessard, M.R. and Smith, L.T.P., "Solar modal structure of the engineering environment," *Proceedings of the IEEE*, 2007, 95 (5), 1085 - 1132.  See also *ESA Space Science News*, "Moving to the rhythm of the Sun," http://www.esa.int/esaSC/SEMJJYUL05F_index_0.html

23) Anders, E. and Grevesse, N., "Abundances of the elements: Meteoritic and solar," *Geochim. Cosmochim. Acta*, 1989, 53, 197-214.

24) Manuel, O.K. and Hwaung, Golden, "Solar abundance of the elements," *Meteoritics*, 1983, 18, 209-222.

25) Manuel, O., Pleess, M., Singh, Y. and Myers, W.A., "Nuclear systematics: Part IV. Neutron-capture cross sections and solar abundance", *J. Radioanalytical and Nuclear Chemistry*, 2005, 266 (2), 159-163. http://www.omatumr.com/abstracts2005/Fk01.pdf

26) Burbidge, E.M., Burbidge, G.R, Fowler, W.A. and Hoyle F. [B2FH], "Synthesis of elements in stars," *Rev. Mod. Phys.*, 1957, 29, 547-650.

27) Harkins, W.D., "The evolution of the elements and the stability of complex atoms," *Journal of the American Chemical Society*, 1917, 39, 856-879.

28) Manuel, O. and Friberg, Stig, "Composition of the solar interior: Information from isotope ratios," *Proceedings of the 2002 SOHO/GONG Conference on Local and Global Helioseismology: The Present and Future*, Big Bear Lake, CA, USA, ESA (European Space Agency) SP-517 (editor: Huguette Lacoste, 2003) pp. 345-348. http://arxiv.org/pdf/astro-ph/0410717

29) Toth, Peter, "Is the Sun a pulsar?" *Nature*, 1977, 270, 159-160.

30) Bohr, Niels " On the constitution of atoms and molecules," *Phil. Mag*., 1913, 26, 1-25.

31) Harutyunian, G.A., "Fragmentation of cosmic objects in the course of their evolution and the possible role of Hubble expansion in this process," *Astrophys.,* 2003, 46, 81-91; *Astrofizika,* 2003, 46 (1), 103-118

32) Manuel, O., Mozina, M. and Ratcliffe, H., "On the cosmic nuclear cycle and the similarity of nuclei and stars," *Journal of Fusion Energy,* 2006, 25, 107-114.

33) Rouse, C.A., "Evidence for a small, high-Z, iron-like solar core," *Astron. Astrophys.*, 1985, 149, 65-72.

34) Kuroda, P.K. and Myers, W.A., "Iodine-129 and plutonium-244 in the early solar system," *Radiochim. Acta*, 1996, 77, 15-20.

35) Boulos, M.S. and Manuel, O.K., "The xenon record of extinct radioactivities in the Earth," *Science*, 1971, 174, 1334-1336.

36) Manuel, O.K., Hennecke, E.W. and Sabu, D.D., "Xenon in carbonaceous chondrites", *Nature*, 1972, 240, 99-101.

37) Sabu, D.D. and Manuel, O.K., "Xenon record of the early solar system", *Nature*, 1976, 262, 28-32.

38) Manuel, O.K. and Sabu, D.D. "Strange xenon, extinct super-heavy elements and the solar neutrino puzzle", *Science*, 1977, 195, 208-209.

39) Ballad, R.V., Oliver, L.L., Downing, R.G. and Manuel, O.K., "Isotopes of tellurium, xenon and krypton in the Allende meteorite retain record of nucleosynthesis", *Nature*, 1979, 277, 615-620.

40) Manuel, O.K. and Sabu, D.D., "Noble gas anomalies and synthesis of the chemical elements", *Meteoritics*, 1980, 15(2), 117-138.

41) Manuel, O., "Origin of elements in the solar system", in *The Origin of Elements in the Solar System: Implications of Post 1957 Observations*, Kluwer Academic/Plenum Publishers, New York, NY, 2000, 589-643.

42) Manuel, O. and Katragada, A., "The Sun's origin and composition: Implications from meteorite studies", in Warmbein, B., ed., *Proceedings of Asteroids, Comets, Meteors (ACM 2002)*, 29 July - 2 Aug 2002, Technical University of Berlin, ESA SP-500, 2003, 787-790.

43) Reynolds, J.H., "Determination of the age of the elements," *Phys Rev. Lett.*, 1960, 4, 8-10.

44) Reynolds, J.H., "Isotopic composition of primordial xenon," *Phys Rev. Lett.*, 1960, 4, 351-354.

45) Kuroda, P.K., "Nuclear fission in the early history of the earth," *Nature*, 1960, 187, 36-38.

46) Fowler, W.A., Greenstein, J.L. and Hoyle, F., "Deuteronomy. Synthesis of deuterons and light nuclei during the early history of the Solar System," *Am. J. Phys.*, 1961, 29, 393-403.

47) Tuli, J.K., *Nuclear Wallet Cards,* 6$^{th}$ ed., Brookhaven National Laboratory, National Nuclear Data Center, Upton, NY, 2000, 74 pp.

48) Manuel, O., Bolon, C., Katragada, A. and Insall, M., "Attraction and repulsion of nucleons: Sources of stellar energy", *Journal of Fusion Energy*, 2001, 19, 93-98.

49) Manuel, O., Miller, E. and Katragada, A., "Neutron repulsion confirmed as energy source", *Journal of Fusion Energy*, 2002, 20, 197-201.

50) Phillips, Tony, "Magnetic portals connect Sun and Earth," *Space Daily*, October 31, 2008. http://www.spacedaily.com/reports/Magnetic_Portals_Connect_Sun_And_Earth_999.html

51) Singer, S. Fred, *Nature, not human activity, rules the climate: Summary for Policymakers of the Report of the Nongovernmental International Panel on Climate Change (NIPCC)*, The Heartland Institute, Chicago, IL 2008, 50 pages.
http://www.sepp.org/publications/NIPCC_final.pdf

52) IPCC, Appendix A to the Principles Governing IPCC Work: *PROCEDURES FOR THE PREPARATION, REVIEW, ACCEPTANCE, ADOPTION, APPROVAL AND PUBLICATION OF IPCC REPORTS*, adopted 19-21 February 2003, page 4/15, Paragraph 4.2: Reports Accepted by Working Groups and Reports Prepared by the Task Force on National Greenhouse Gas Inventories: http://www.climatescience.gov/Library/ipcc/app-a.pdf

53) IPCC Reports, *IPCC Fourth Assessment Report*
http://www.ipcc.ch/ipccreports/ar4-syr.htm

54) McLean, John, "Peer Review? What Peer Review? Failures of scrutiny in the UN's Fourth Assessment Report," Science & Public Policy Institute, Washington, DC, 2007, 30 pp. http://icecap.us/images/uploads/McLeanIPCCReviewfinal9-5-07.pdf

55) McKitrick, R. et al. 2007. *Independent Summary for Policymakers IPCC Fourth Assessment Report*, Fraser Institute
http://www.fraserinstitute.org/Commerce.Web/product_files/Independent Summary5.pdf

56) Eisenhower, Dwight D., "Eisenhower's Farewell Address to the Nation, January 17, 1961, http://mcadams.posc.mu.edu/ike.htm